\documentclass[twocolumn,floatfix,secnumarabic,amssymb, nobibnotes, aps, prd]{revtex4-1}

\usepackage{amsmath,amssymb}

\usepackage{nameref,hyperref}
\usepackage{graphicx}

\usepackage{comment}

\DeclareMathAlphabet{\pazocal}{OMS}{zplm}{m}{n}
\usepackage{bm}

\begin{document}

\title{Impact of temporal correlations on high risk outbreaks of independent and cooperative SIR dynamics}

\author{Sina Sajjadi$^{1}$} 
\author{Mohammad Reza Ejtehadi$^{1}$}
\author{Fakhteh Ghanbarnejad$^{1}$}

\email[Corresponding author: ]{fakhteh.ghanbarnejad@gmail.com}
\affiliation{$^1$Department of Physics, Sharif University of Technology, Tehran, Iran}

\begin{abstract}
        We first propose a quantitative approach to detect high risk outbreaks of independent and coinfective SIR dynamics on three empirical networks: a school, a conference and a hospital contact network. This measurement is based on the k-means clustering method and identifies \textit{proper samples} for calculating the \textit{mean outbreak size} and \textit{the outbreak probability}. Then we systematically study the impact of different temporal correlations on high risk outbreaks over the original and differently shuffled counterparts of each network. We observe that, on the one hand, in the coinfection process, randomization of the sequence of the events increases the mean outbreak size of high risk cases. On the other hand, these correlations don’t have a consistent effect on the independent infection dynamics, and can either decrease or increase this mean. While randomization of the daily pattern correlations has no significant effect on the size of outbreak in either of the coinfection or independent spreading cases. We also observer that an increase in the mean outbreak size doesn't always coincide with an increase in the outbreak probability; therefore we argue that merely considering the mean outbreak size of \textit{all realizations} may lead us into misestimating the outbreak risks. Our results suggest that some sort of randomizing contacts in organization level of schools, events or hospitals might help to suppress the spreading dynamics while the risk of an outbreak is high.
\end{abstract}
\maketitle

\section*{Introduction}
Infectious diseases have had drastic impacts on human health throughout the history, resulting in major social and economical disruptions \cite{hays_2006}. Mathematical models are well known methods, used in order to achieve better understanding and prediction of this phenomena \cite{anderson_may_2010, keeling2011modeling}. Susceptible-Infectious-Recovered (SIR) model \cite{kermack_mckendrick_1927} is one of the most basic and common models for describing and predicting the epidemics of the contagious diseases. This model and its variations have been developed to model patterns of spreading dynamics in different scenarios. For instance, some models discuss how considering contact networks of the host population can alter dynamics, e.g. epidemic threshold \cite{newman2002spread, dorogovtsev2008critical, newman2006structure, newman_2018, pastor2001epidemic, pastor2001epidemic_2, pastor2002epidemic, moreno2002epidemic}. Some other works improved models by considering temporality of the contacts \cite{speidel2017epidemic, masuda2017temporal,rocha2013bursts} and some studies focus on impact of some temporal correlations on spreading dynamics \cite{fakhteh-hospital, smallbutslowworld}. Moreover, some models consider the case of coinfective diseases: when getting infected by one disease, alters the chance of getting infected by another one. They studied cooperative or competitive spreading dynamics in the mean field approximations and on complex networks with different topologies \cite{fakhteh-epl, fakhteh-nature, grassberger2016phase, chen2017fundamental, rodriguez2019particle, goel2018modelling, sanz2014dynamics, pinotti2019interplay, soriano2019markovian, zarei2019exact}. 

Despite all these successes, we lack a quantitative method to measure systematically how temporal correlations of the contact networks affect the spreading dynamics, specially when two or more dynamics interact. On the other hand, the increasing amount of empirical contact data, better computational performance, and our interest in understanding real-world situations, lead us to propose a quantitative measurement for studying the impact of temporal correlations of the empirical networks on independent and cooperative SIR dynamics. To address these issues, we run the independent and coupled SIR model on three empirical temporal networks and their randomized counterparts. Our proposed measurement is based on the k-means clustering \cite{wikipedia_2020_kmeans} and determines which samples to pick up and average in order to detect the impact of different correlations on outbreak size and probability of the outbreak, theses samples represent the high risk outbreaks.

\section*{Materials and methods}
\subsection*{Topology: Empirical Temporal Networks}\label{temporal-network-data}

Here we study three different empirical temporal networks. All data sets are representative of the interactions among individuals, using wearable sensors which detect close-range contacts between individuals. These data sets contain the list of contacts recorded within a specific time period. Every contact is characterized by the labels of the individuals conducting the interaction and the time of the interaction; their format informs the contact sequence picture which contrasts with the interval graph \cite{temporalnetworksholme,MRRM}. The networks are:

\begin{enumerate}
	\item \textit{Hospital Network}: "Contacts between patients and health-care workers in a hospital ward in Lyon, France, from Monday, December 6, 2010 at 1:00 pm to Friday, December 10, 2010 at 2:00 pm"	\cite{hospitalnet-2}.
	\item \textit{Conference Network}: "Face-to-face interactions between ACM Hypertext 2009 conference attendees"\cite{conferencenet-2}.
	\item \textit{Primary School Network}: "Contacts between the children and teachers in a primary school in Lyon, France during two days in October 2009"\cite{primaryschoolnet-2, primaryschoolnet-3}.
	
\end{enumerate}

Some characteristics of these networks are summarized in the table \ref{tbl:data_characteristics}.

\begin{table*}

	\centering
	\begin{tabular}{|c|c|c|c|c|c|c|c|}
	\hline

		Data & Vertices \# & Contacts \# & Duration & Resolution & link/t Ave\\
		\hline

		Hospital & 75 & 32424 & 4 days & 20s & 0.093\\
		Conference& 113 & 20818 & 3 days & 20s & 0.098\\
		Primary School& 242 & 125773 & 32 hours & 20s & 1.076\\
		
		\hline
	\end{tabular}
	\caption{Some characteristics of three different empirical temporal networks: Hospital Network\cite{hospitalnet-2}, Conference Network \cite{conferencenet-2} and Primary School Network \cite{primaryschoolnet-2}}
	\label{tbl:data_characteristics}

\end{table*}

\subsubsection*{Shuffling temporal correlations}
In order to study effects of different temporal correlations on any spreading phenomena, one can shuffle the correlations. Any shuffling method randomizes some correlations and preserves the rest. Comparison of the spreading dynamics on the original and shuffled networks determines the impact of that specific temporal correlation on the dynamics. 
\\
Let's first introduce some attributes of our temporal networks:
\begin{enumerate}
	\item D: Daily pattern: as is it's shown in the first set of graphs, daily patterns are the frequency of events, namely contact occurrences.
	\item C: Community structure.
	\item B: Bursty event dynamics of single links.
	\item W: Weight-topology correlations.
	\item E: Event-event correlations between links.
\end{enumerate}
Some of the shuffling methods we practiced are titled in \cite{smallbutslowworld} by the network attributes they preserve, also a comprehensive naming convention has been designed in \cite{MRRM} to label different methods of temporal network shufflings; we will introduce our shuffles, using both conventions.

We define the time-stamps which every edge has appeared as the “single-link event sequence”, and the number of appearances of an edge as weight.

The shuffling methods, also demonstrated in Fig. \ref{shuffles} are listed bellow:

\begin{figure*}[]
	\begin{center}

		\includegraphics[width=\linewidth]{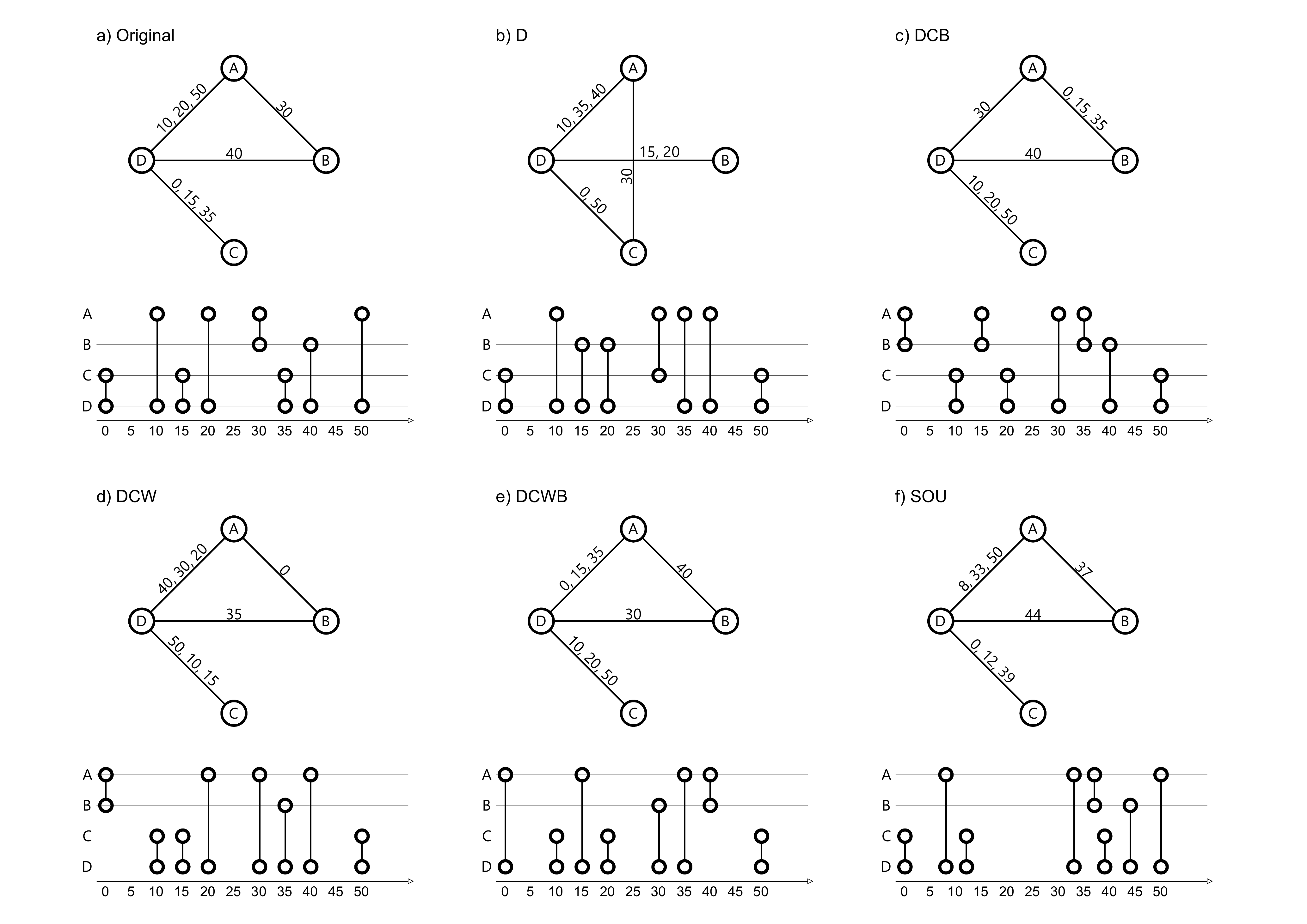}
		\caption{Visualization of a sample original network and its shuffled counterparts. For each network, the upper figure demonstrates the aggregated network with times of contacts as labels on edges. And the lower figure demonstrates a timeline for vertices, while the vertical line denotes the existence of a contact between two vertices at that specific time-step.}
		\label{shuffles}
	\end{center}
\end{figure*}

\begin{enumerate}
	
	\item DCWB (equal-weight link-sequence shuffled): "Whole single-link event sequences are randomly exchanged between links having the same number of events. Temporal correlations between links are destroyed"\cite{smallbutslowworld}, titled $P[\bm w, p_\pazocal{L}(\bm \Theta)]$ in \cite{MRRM}.\\
	
	\item DCB (link-sequence shuffled): "Whole single-link event sequences are randomly exchanged between randomly chosen links. Event-event and weight-topology correlations are destroyed"\cite{smallbutslowworld}, titled $P[\pazocal{L}, p_\pazocal{L}(\bm \Theta)]$ in \cite{MRRM}.\\
	
	\item DCW (time-shuffled): "Time stamps of the whole original event sequence are randomly reshuffled. Temporal correlations are destroyed"\cite{smallbutslowworld}, titled $P[\bm w, \bm t]$ in \cite{MRRM}.\\
	
	\item D (configuration model): "The original aggregated network is rewired according to the configuration model, where the degree distribution of the nodes and contentedness are maintained but the topology is uncorrelated. Then, original single-link event sequences are randomly placed on the links, and time shuffling as above is performed. All correlations except seasonalities like the daily cycle are destroyed"\cite{smallbutslowworld}, titled $P[\bm k, \mathbb{I}_\lambda, p( \bm w ), \bm t]$ in \cite{MRRM}. The only preserved attribute is the daily patterns, since we use the previous time-stamp to randomly shuffle our new edges on them. Since this kind of shuffling may destroy the connectivity of our network, if so one needs to use the giant cluster of the produced network.
	
	\item SOU (same-ordered): We introduce a new method of shuffling for temporal networks. Using a uniform random distribution, we create a new time-series for our temporal network, and we assign the new time-series to the occurrence time of our contacts, while preserving the ordering of the events.
	Unlike the previously implemented shuffles, the ordering of the events (appearance of the edges) will remain intact, but the time difference correlations (daily patterns) will be destroyed.
	
\end{enumerate}

Please note that all of the three original aggregated networks have a single connected component as big as the system size. Also all of the shuffling methods preserve the size of connected component for the aggregated shuffled network.

\subsection*{Dynamics}
\begin{figure}[]
	\begin{center}
		\includegraphics[width=\linewidth]{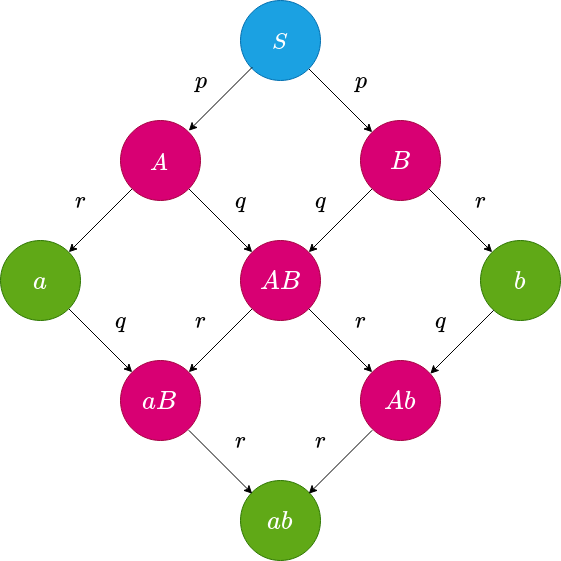}
		\caption{Different states in the SIR-SIR model \cite{fakhteh-epl} and the probabilities of switching between states.
			Capital letters denote the infectious state regarding one disease, small case letters denote the recovered states. Each arrow indicates a transition with a certain probability. State S illustrated by color blue, denotes the agents' initial states. Blocks illustrated by color pink and green, respectively represent infectious and recovered states.
		}
		\label{9states}
	\end{center}
	\end{figure}

In this model every agent can be in any of three different states: "S-I-R" (Susceptible-Infected-Recovered) based on its status regarding any of the two disease. This in total makes up 9 different states which are demonstrated in Fig. \ref{9states}. The simulations occur using the rejection-based modeling \cite{rejection-based}.
on each time step,

\begin{enumerate}
	\item every infectious agent, turns its susceptible neighbors to an infectious agent for the same disease with probability $p$.
	\item every infectious agent, infects its neighbors which have been infected (recovered) by the other disease with probability $q$ with the secondary infection.
	\item every infectious agent recovers from each disease by the probability $r$
	\cite{barrat}.
	
\end{enumerate}

\begin{figure*}

	\centering
	{\includegraphics[width=.3\linewidth]{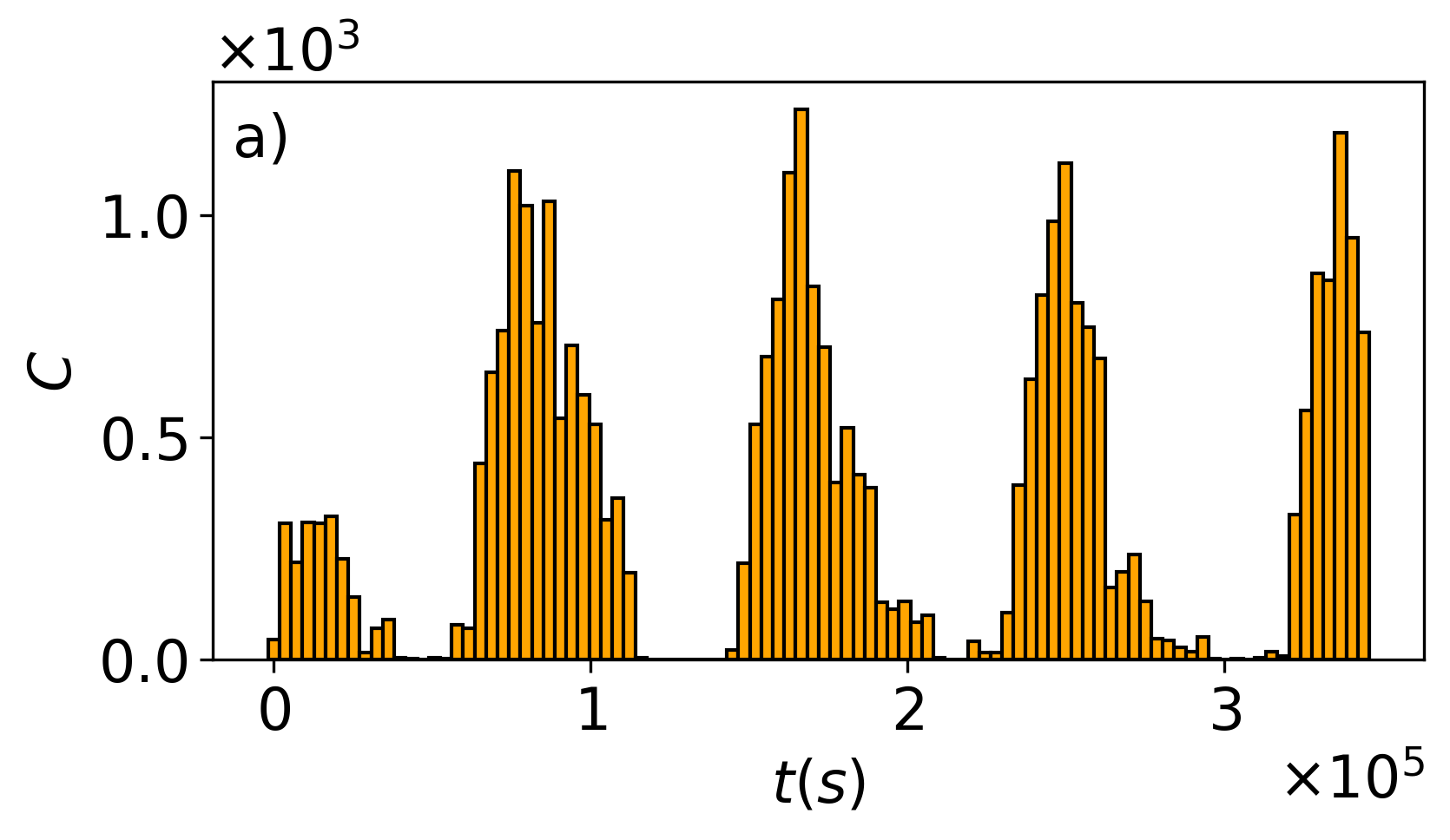}}
	{\includegraphics[width=.3\linewidth]{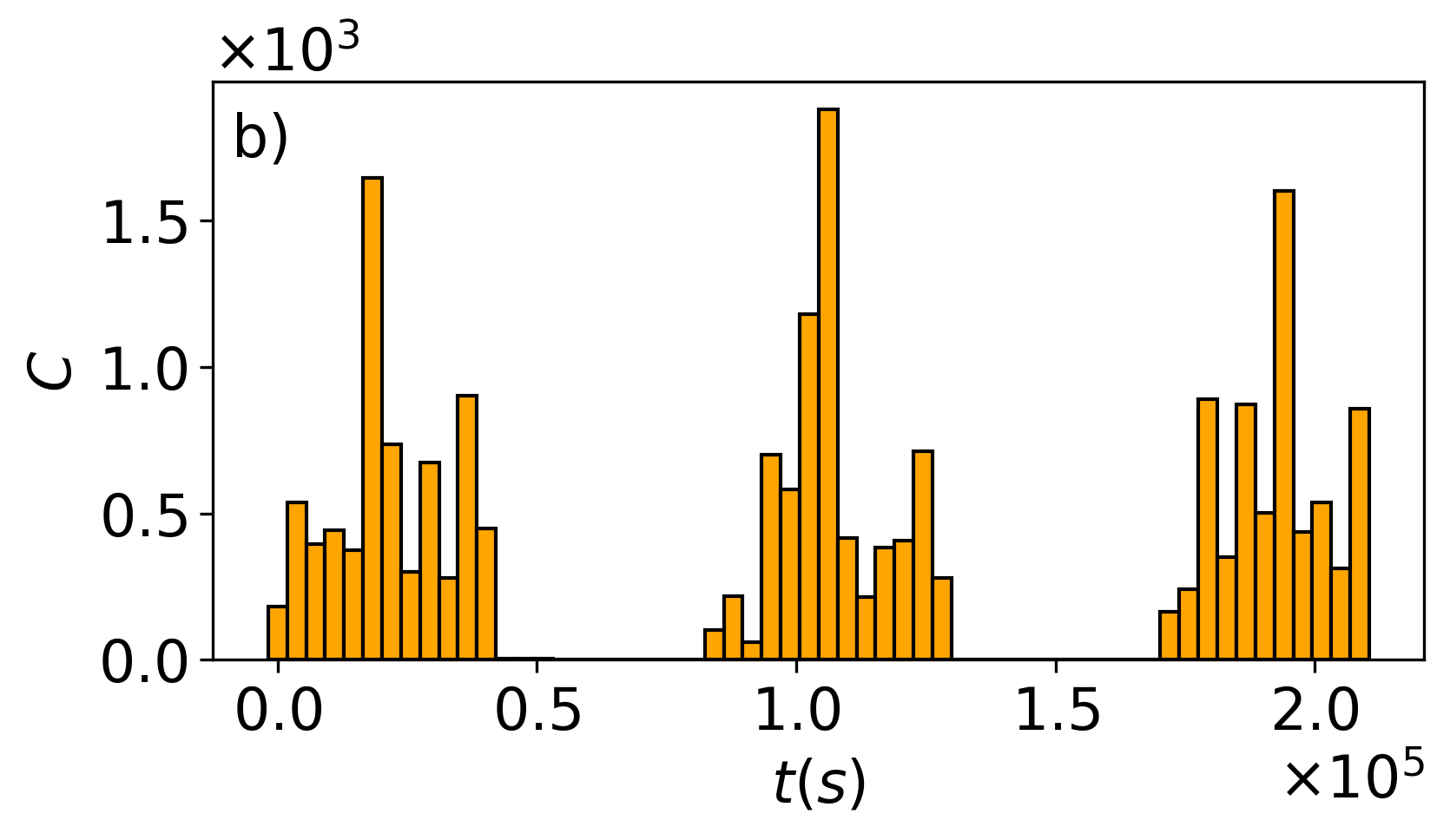}}
	{\includegraphics[width=.3\linewidth]{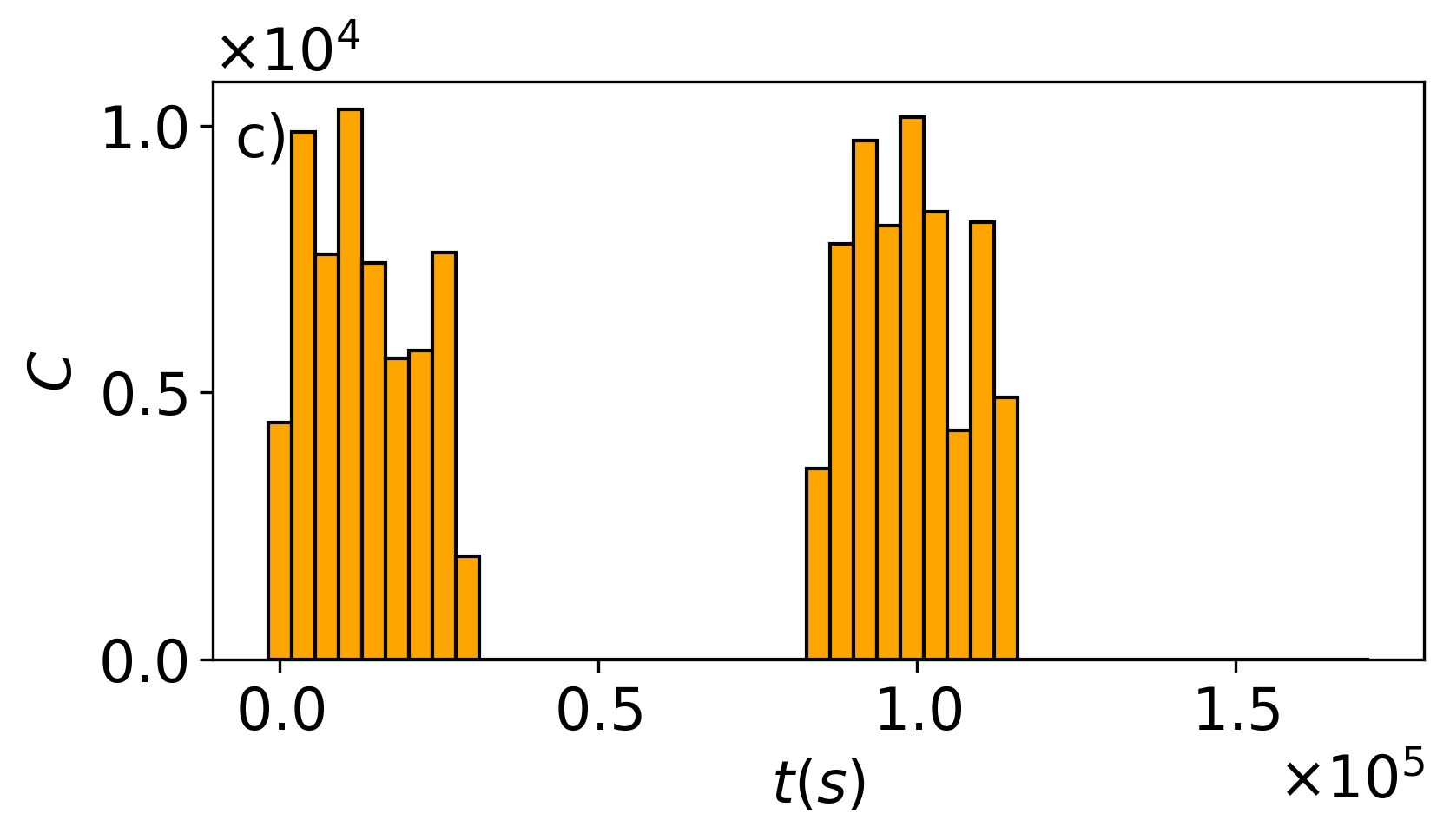}}
	
	\caption{The periodic behavior of the number of contacts for 3600s aggregated time intervals for the a) Hospital Network, b) Conference Network and c) Primary School Network.}
	\label{Fig:edges-hist}
\end{figure*}

The two diseases generally act independently, except that when an agent has been infected with one disease (whether it's recovered or still infectious), the probability for becoming infected by the other disease $q$. The three parameters $r$, $p$ and $q$ are our control parameters. Since the number of control parameters are relatively high, we set a specific value for $r$, for each temporal network, considering the frequency and distribution of its contacts, see Fig. \ref{Fig:edges-hist} and Table \ref{tbl:data_characteristics}. As it is noted in \cite{fakhteh-hospital} we notice the inactivity periods within each activity histogram. To examine the effects of each network's dynamics, we need to consider a recovery rate, low enough, which enables the infection to survive during the inactivity period. On the other hand, considering a very low recovery rate will decrease the speed of the spreading dynamic. It has been reported that in cases in which, the speed of the dynamic on the network (spreading) and the speed of the dynamic of the temporal network have a large difference, the effect of the spreading over the temporal will be the same as on its weighted aggregated counterpart \cite{dreview}. Therefore, to capture the temporal effects, we consider the length of the valleys as an indicator of the suitable recovery rate, please see Fig. \ref{Fig:edges-hist}. by considering the time-scale of valley's for both hospital and conference network which is approximately the size of 1000 steps(each step is 20 seconds), we assign $r=0.001$. In the case of primary school the size of the valley is of the order of 5000 time steps (each step is 20 seconds) so we assign $r=0.0002$. We also consider two cases for $q$, called cooperative and independent spreading as indicated in \cite{fakhteh-hospital}. In the former situation, we set $q=1$ so the probability of acquiring a second disease would be higher than the first one (the range of values for $p$ are drastically lower than 1). And in the latter, we set $q=p$, so the spreading would be totally independent.\\
The simulations run until the dynamics reach the stationary states, i.e. all of the agents will be in one of the blue or green states of Fig. \ref{9states}, while a temporal-periodical boundary condition is applied to the contact network.\\
Our initial condition is set to a single randomly chosen doubly infected node.
\subsection*{Macroscopic Observables (Order Parameters)}
In such dynamics, the most common macroscopic observable is the average fraction of infected individuals, in this case $<ab>$. However, as shown in \cite{fakhteh-nature} and \cite{fakhteh-hospital} due to the branching effect of the coinfection dynamics, average won't be a good indicator of suchh epidemic behaviors. Moreover, this may not be specific to the situation of coinfection and due to finite size effect one can observe branches with somewhat similar results. The histograms in Fig. \ref{fig:KMean-Measure}, left panel and also in the supplementary material, show at least two branches, one formed around the 0 value (which may be cut out in our presentations) and the other, i.e. outbreak branch, formed around a higher value. This outbreak branch represents the high risk outbreak instances.
Therefore instead of averaging the whole distribution, we only average over the outbreak branch, namely mean out break size ($\overline{ab}$), and also look at another order parameter, namely outbreak probability ($P_{ab}$):  the probability for a realization to land on the outbreak branch. In simple terms the first parameter indicates the pervasiveness of an outbreak, and the second implies how likely it is that an outbreak would occur. These parameters are mathematically defined in equations \ref{prob-equation} and \ref{mean-equation}, where $\rho_{ab}$ is the fraction of doubly infected agents, $\pi ( \rho_{ab}(p*) )$ is the distribution of $\rho_{ab}$ for a specific $p*$ value of the control parameter and integration over outbreak branch ($OB$), accounting for high risk outbreaks. 

\begin{equation} 
P_{ab}(p*) = \int_{OB}^{ } \pi ( \rho_{ab}(p*) ) d  \rho_{ab}(p*)
\label{prob-equation}
\end{equation}
\begin{equation}
\overline{ab} = \int_{OB}^{ } \pi ( \rho_{ab}(p*) ) \rho_{ab}(p*) d ( \rho_{ab}(p*) )
\label{mean-equation}
\end{equation}

Now a problem may arise that the main two branches are not well distinguishable, specially in the case of coinfections on \cite{fakhteh-nature}. So we introduce a new method, using k-means clustering
\cite{wikipedia_2020_kmeans}
to determine their values.

\subsection*{k-means Clustering}
\begin{figure*}
	\centering
	\includegraphics[width=\linewidth]{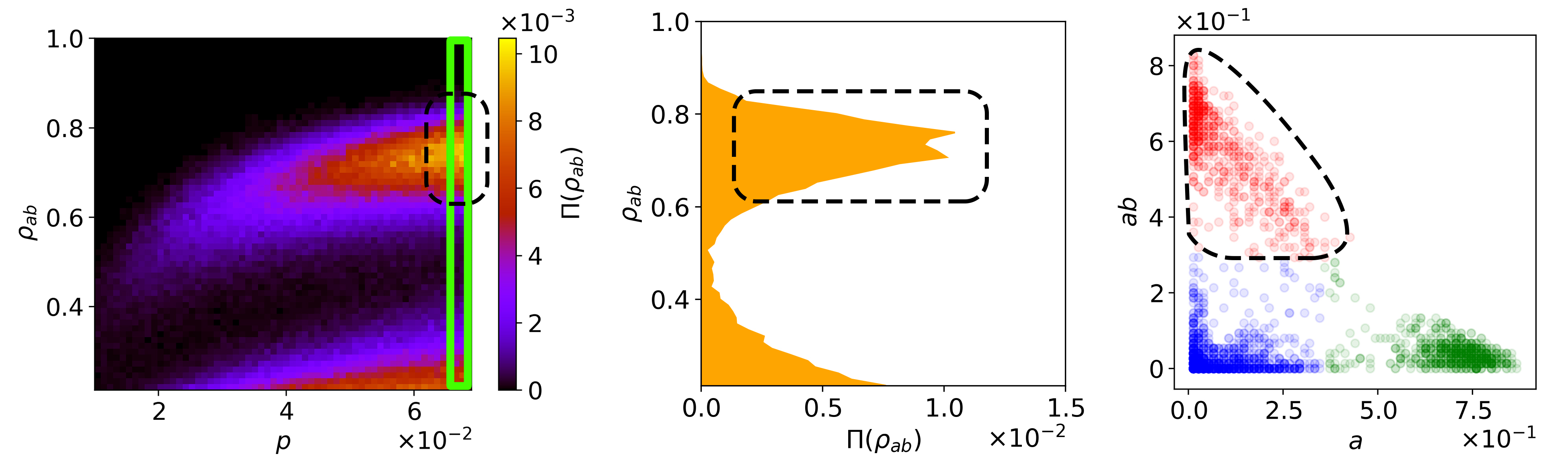}‎
	\caption{Systematic measurement of $P_{ab}$ (Eq. \ref{prob-equation}) and $\overline{ab}$ (Eq. \ref{mean-equation}). This figure demonstrates, how the k-means clustering measure works, for example here for coinfection dynamics on the DCWB shuffled hospital network while $q = 1$ and $r = 0.001$. The left panel shows the density of the ${ab}$ population, and the probability that a realization ends to the given value of the density ($P_{ab}$: color code) while varying p, the first infection probability. The middle panel distinguishes precisely the two epidemic branches in the left panel at $p=0.069$, the vertical window. The right panel shows the fraction of individuals infected by disease a (axis x) and by both diseases (axis y), within each realization. Each point denotes a single realization and different colors indicate different clusters: red (doubly infected outbreaks), green (single infected), blue (no outbreaks). The lower parts of left and middle panels are cut out in order to better emphasize on discrepancies in higher $ab$ values.
	The dashed shapes, encircle the realizations which make up the outbreak branch (OB) in each illustration. The number of realizations is 50000, but for illustrative purposes in the right panel, only a sample of 5000 realizations are depicted.}
	\label{fig:KMean-Measure}
\end{figure*}

Dynamics evolves following one of these scenarios: 1- No considerable growth for either of the diseases. It means that the dynamics dies out early in the process and lands on the lower branch in Fig. \ref{fig:KMean-Measure}, left and middle panels. 2- One disease dies out early and the other disease gaining a considerable growth independently. It means that cooperation does not affect the process. Similarly, also in this scenario, dynamics lands on the lower branch. 3- Both diseases gaining a considerable growth. This type of dynamics ends up on the upper branch in Fig. \ref{fig:KMean-Measure}, left and middle panels.\\
Now to address the issue of distinguishing the outbreak branch from the lower branch, we look at not only the $ab$ infected ones, but also the $a$ (or $b$ as the dynamics is symmetrical) infected ones.
Fig. \ref{fig:KMean-Measure}, right panel depicts these scenarios in 3 clusters: blue, green and red respectively.
k-means clustering method provides us a systematic measure to classify these cases and find the red cluster, right panel, as counterpart of the upper branch in the left and middle panels.
k-means clustering is a method of partitioning data into a specific number of clusters so that data points within the same cluster will have less distance from each other, compared to data points of other clusters. Here $a$ and $ab$ are the parameters employed to devise an euclidean distance between the data points and as we discussed, we set the number of requested clusters for the k-means algorithm to 3.  

By defining the red cluster which corresponds to the outbreak branch, we can now quantitatively define $P_{ab}$ as the fraction of realizations which fall in the outbreak branch and $\overline{ab}$ as the outbreak branch center.

\section*{Results}
We proceed to span the phase space of $p$ to observe the behavior of the number of doubly infected agents ($ab$). We first derive the distribution of the $ab$ size for each set of parameters and networks. To visualize our results we use 2D histograms consisting of 1D histograms for each $p$ value, as shown in Fig. \ref{fig:KMean-Measure}, left panel and also in the supplementary material.
\\

The first point we observe by investigating \ref{fig:KMean-Measure}, left panel and all of the figures in the supplementary material is the change in $p_{c}$ or the outbreak threshold. This parameter which indicates the lowest value of $p$ that causes a significant outbreak, is lower for all coinfection spreadings ($q=1$), compared to their independent ($q=p$) spreading counterparts.
\\

\begin{figure*}
	\begin{center}
		\includegraphics[width=1.0\linewidth]{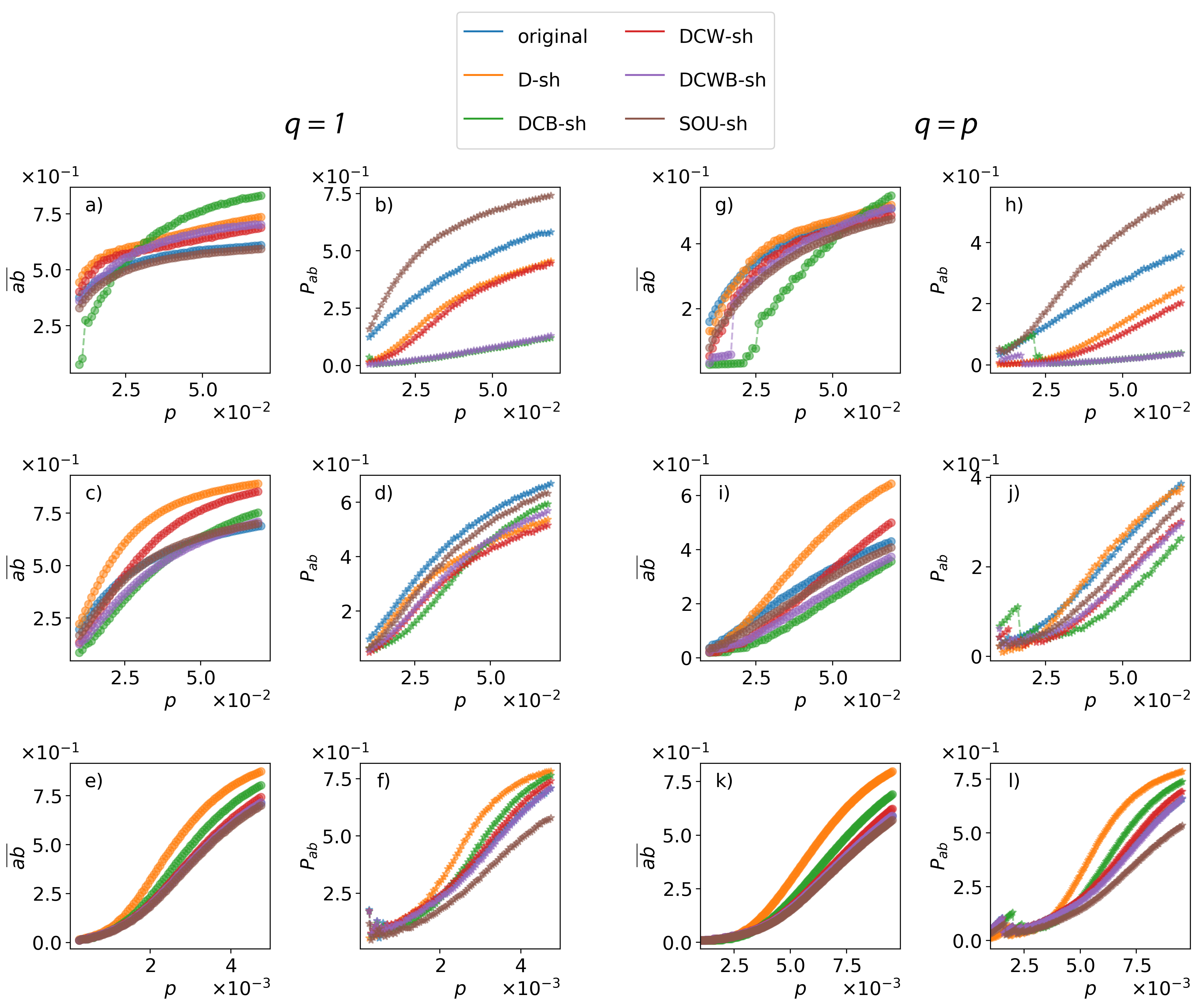}
		\caption{Mean of the outbreak size ($\overline{ab}$), columns 1 and 3, and the outbreak probability ($P_{ab}$), columns 2 and 4, for coinfective SIR-SIR ($q=1$), columns 1 and 2, and independent SIR-SIR dynamics ($q=p$), columns 3 and 4. First, second and third rows shows results for hospital, conference and primary school networks respectively. These values are respectively obtained based on the introduced method in the Fig. \ref{fig:KMean-Measure} right panel, i.e. the percentage and average $ab$ of red points in the right panel.}
		\label{fig:orderparameters}
	\end{center}
\end{figure*}

Furthermore, we calculate $P_{ab}$ and $\overline{ab}$ for different shuffled versions of our networks in Fig. \ref{fig:orderparameters}.

As mentioned, all of the shufflings starting with capital D (we will call them D family shuffles), retain the daily patterns i.e. original time-series of the network, while destroying the ordering of the events, (event-event correlations) and therefore causality between the events; whereas SOU shuffling retains the event-event correlations and randomizes the time-series.\\

We firstly summarize our observations for \textbf{coinfection}, Fig. \ref{fig:orderparameters} panels a-f, scenarios: 

\begin{enumerate}
	\item D family shuffles will increase the value of $\overline{ab}$ for high $p$ values, therefore in case an outbreak happens it would be more hazardous. This may be due to the fact that in the empirical networks with an underlying spatio-temporal structure, there exist a localized behavior, meaning that individuals will interact with the same group of people instead of exploring new people, therefore they keep the disease in close neighborhood during short periods of time. Additionally spatio-temporal structures will raise transitive relations, so the individual will have stronger clustering compared to randomized networks. For example, if an individual "Sina" is interacting with another individual "Fakhteh", and "Fakhteh" is interacting with "Reza", then the probability that "Sina" and "Reza" interact with each other in a short period of time, is higher than the average probability of interaction between two individuals, this is due to both spatio-temporal features of the network and the social relations among individuals. By destroying the event-event correlations the individuals will have a higher chance to interact with more people during a short period, therefore it leads to an increase in the size of a possible outbreak. Due to the stronger clustering in original networks, the pathogens can be trapped in an spatio-temporal structure. Since D family shuffling methods randomize these clusters, the trapping probability get reduced and two diseases can interact more often. Thus if they could meet, the fraction of doubly infected people would increase.
	
	\item SOU shuffle has no significant effect on $\overline{ab}$. This agrees with the results in \cite{linkbirthanddeath}, albeit there only a single SIR dynamics is studied. Holme and Liljeros have performed a type of shuffling named Inter-event Interval Neutralized (IIN) method, which sets a uniform time distribution to the activity of each edge, within its original first (birth) and last (death) appearance. By considering the average number of infected individuals during SIR spreading, they concluded that inter-event time distributions don't have a significant effect on the spreading, while the time of birth and death of a link are important. One should note that, SOU shuffling keeps the order of the events and redistributes the time intervals between all events while IIN method \cite{linkbirthanddeath} redistributes the time intervals between the events of the same type. This means that IIN may lead to reordering of the events in small time durations; but these methods are of the same nature since they both keep the order of appearance of the edges and the activity clock \cite{burstyhuman} between birth and dead of a link intact, but they destroy the inter-event time distributions.

	However, as we can observe in Fig. \ref{fig:orderparameters}, panels b, d and f, SOU shuffles can either decrease or increase the outbreak probability ($P_{ab}$). This means that inter-event time distributions may affect the dynamics but we need to look at proper order parameters and proper
	averaging. For instance, Holme and Liljeros have only studied the outbreak size ($\overline{ab}$) averaged over all realizations and concluded that inter-event time distributions have no significant impact on the dynamics.

	\item Some correlations which hinder the process of coinfection for a range of control parameters, may also enhance the spreading for another set of parameters. Fig. \ref{fig:orderparameters} panels a and c, present such examples, where the DCB shuffles have less $\overline{ab}$ in comparison to the original networks for $p<0.03$ (hospital) and $p<0.05$ (conference). Nevertheless, for the DCB shuffles, $\overline{ab}$ manages to surpass the value of the original network for greater values of the control parameter. Hence it should be noted that the effect of link weight correlations on spreading phenomena is highly dependent on the range of the control parameter. 
	
	While recently some studies discussed that temporal correlations can either facilitate the spreading dynamics or weaken such processes\cite{dreview};
	Here, our results show that both scenarios can happen depending on where the system is in the parameter space. Which represents the dynamical and topological characteristic-times of the system.

\end{enumerate}
Secondly, we compare the results of coinfection ($q=1$) Fig. \ref{fig:orderparameters} panels a-f with independent spreading dynamics ($q=p$) Fig. \ref{fig:orderparameters} panels g-l we observe that:
\begin{enumerate}
	
	\item
	Both order parameters $\overline{ab}$ and $P_{ab}$ are less in comparison to the coinfection case ($q=1$).
	\item 
	D family shuffles don't show a consistent effect on $\overline{ab}$, while they increase this value for coinfections. This signifies that the order of the events and also spatio-temporal correlations have a more intensive effect on coinfection compared to the independent spreading dynamics.
	This could be caused by the fact that formation of the spatio-temporal communities may constraint the spreading of each infection in separate communities; considering that a collision of the two diseases, leads to a greater effect on the $\overline{ab}$ for the coinfection, this type of correlations have a greater effect on the coinfective dynamics.
	In another word, neighboring effects caused by the existence of smaller communities in temporal networks, decrease the chance of two diseases interacting with each other, so lack of interaction between two diseases for proper time can impact stronger the coinfection dynamics compared to the independent dynamics.
	
	\item
	SOU shuffle doesn't have any significant effects on $\overline{ab}$, though it can affect (increase or decrease) $P_{ab}$, same as the situation with coinfection.
	\item 
	Similar to the coinfection cases, Fig. \ref{fig:orderparameters}g shows that some correlations which hinder the process of spreading in a range of parameter $p$, may also enhance it in other ranges.		
\end{enumerate}

\begin{figure}
	‎\centering‎
	‎\includegraphics[width=\linewidth]{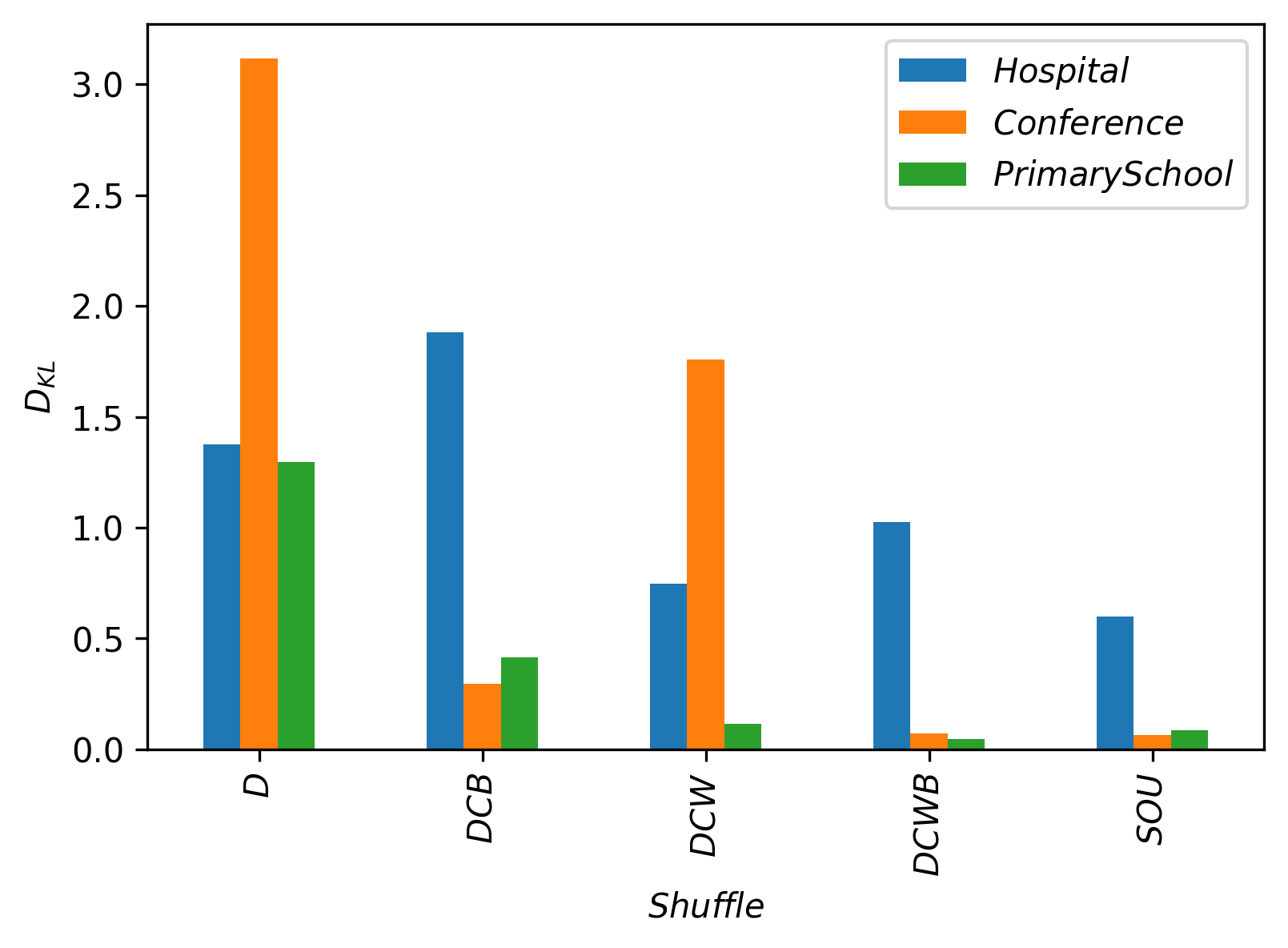}‎
	‎\centering‎
	\caption{
		The value of cross-entropy for the dynamic results under each shuffle in relation to the original network for the value: 
		$p = 0.07$ for hospital and conference networks and $p=0.005$ for the primary school network.
		This parameter indicates the difference between the result of $ab$ distribution for coinfection ($q=1$) on each shuffled network and its original counterpart.
	}
	\label{crossentropy}
	‎\end{figure}

Finally as an alternative we can quantitatively compare the results of the dynamics under different shuffles, we compute the cross entropy or Kullback–Leibler divergence \cite{Kullback} for the probability distribution of $ab$ for each shuffled network in relation to the original network.
To avoid the singularity caused by the logarithm in the cross entropy formula we increase the values of the histogram by the minimum non-zero value and then normalize the probability distribution again. The values for cross entropy with the original network for our shuffled networks can be seen in Fig. \ref{crossentropy}. We observe that generally the dynamics have the closest behavior on the SOU shuffled networks in comparison to the original networks. Since the SOU shuffling keeps the order of events intact, we can conclude that for such dynamics the sequence of events has the most significant effect on the results. Furthermore, we show that although Kullback–Leibler divergence is taking to the account the whole $ab$ distribution (Fig. \ref{fig:KMean-Measure} middle panel) in comparison to the K-means clustering which focuses only on the outbreak branch (Fig. \ref{fig:KMean-Measure} red dots in the right panel), the two measures agree that D shuffle has the greatest impact on the conference and the primary school networks Fig. \ref{fig:orderparameters} panels c, e, while DCB has the greatest impact on the hospital network Fig. \ref{fig:orderparameters} panel a.
‎‎

\section*{Discussion}
In summary, in this work, we studied the effects of various temporal correlations on the spreading process, both independent infection and coinfection. The dynamics were SIR type and simulated on three different temporal empirical networks, as well as their several shuffled counterparts. Each of the shuffling methods preserved some temporal correlations and randomized others. For instance, we introduced SOU shuffling which keeps the ordering of the events intact while randomizing the frequency of the events. We argued that in order to see the impact of any temporal correlation we need to investigate properly two order parameters: the probability of outbreak ($P_{ab}$) and the outbreak size ($\overline{ab}$) as macroscopic observables. Moreover, since in so many cases a simple ensemble averaging may lead to misinterpretation, therefore, we introduced a systematic measurement to identify the proper samples which represent the high risk outbreaks, for calculating these two order parameters. This measurement is based on the k-means clustering method. Furthermore, we introduced another alternative method, namely Kullback–Leibler divergence which calculates the difference between distributions of $ab$ for shuffled and original networks. While k-means clustering identifies the proper samples and then takes an average over them, Kullback–Leibler divergence computes the discrepancy between two histogram cell by cell and then takes an average; We showed that both measurements, agree on which shuffling has dominant impact on which network. 

We showed that cooperation between two diseases facilitate the spreading dynamics on original networks as well as shuffled networks. In the coinfection process, randomization of the sequence of the events makes the outbreak more pervasive, i.e. $\overline{ab}$ decreases. On the other hand, these correlations don't have a consistent effect on the independent infection dynamics, and can either decrease or increase the outbreak size. This point indicates that the ordering of the events and spatio-temporal features have a greater effect on coinfection. In both independent infection and coinfection, daily patterns have no significant effect on the outbreak size, but they can change the probability of outbreaks for various networks.
The last but not the least, in order to understand the impact of temporal correlations on spreading phenomena, not only the proper order parameter and proper averaging matters, but also the dependency on the range of the control parameter matters.

Our proposed systematic measurement provides a more precise method to trace the macroscopic observables. Thus it can help us to improve the epidemic risk calculations \cite{fakhteh-hospital}. Also our results can help organizers and managers to better organize meetings and institutes, in order to decrease high risk outbreaks. For instance, some sort of shuffling of the shift charts in a hospital can decrease the probability of a high risk outbreak.

This method not only improves our understanding of the dynamics "on" the networks, but also can open a road to understand better the topological features of the temporal empirical networks, dynamics "of" the networks.

\section*{Code Availability Statement}
The simulation engine written in C++, and analyzing tools written in Python are parts of the Epyc package, developed by SS, and are available under GPLv3, at \url{https://github.com/Sepante/Epyc}.

\section*{Data Availability Statement}
The empirical network datasets analyzed for this study are distributed to the public under a Creative Commons Attribution-NonCommercial-ShareAlike 3.0 Unported (CC BY-NC-SA 3.0) license; and are available on \textit{sociopatterns.org}.


\section*{Acknowledgments}
Authors thank Fariba Karimi for her helpful discussions and Arash Ahmadian for designing Fig. \ref{shuffles}. F.Gh. acknowledges partial support by Deutsche Forschungsgemeinschaft (DFG) under the grant (idonate project: 345463468). 

\section*{Supporting Material}

\begin{figure*}[]
	\begin{center}
		\includegraphics[width=\linewidth]{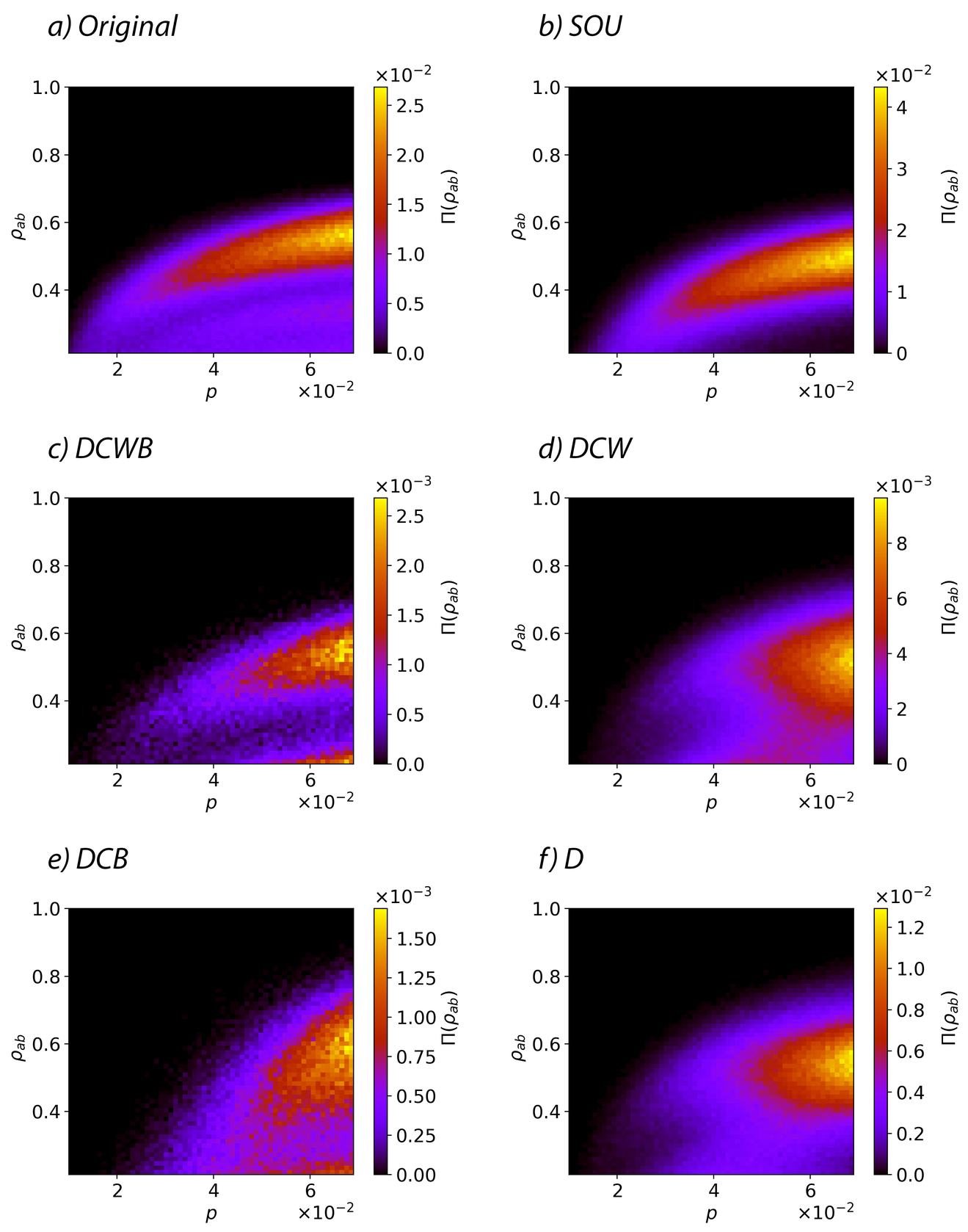}
		\caption{
			The results of \textbf{independent SIR-SIR infection ($\mathbf{q=p}$) simulation on the hospital network}, the x axis is the control parameter $p$, the y axis is the size of the final doubly infected agents ($ab$), and the color axis denotes the percentage of realization with the specific value of $ab$.
			Please note that the coloring scale is different on each graph, to clarify the discrepancy between different regions of each graph.
		}
		\label{hospital-nc}
	\end{center}
\end{figure*}

\begin{figure*}[]
	\begin{center}
		\includegraphics[width=\linewidth]{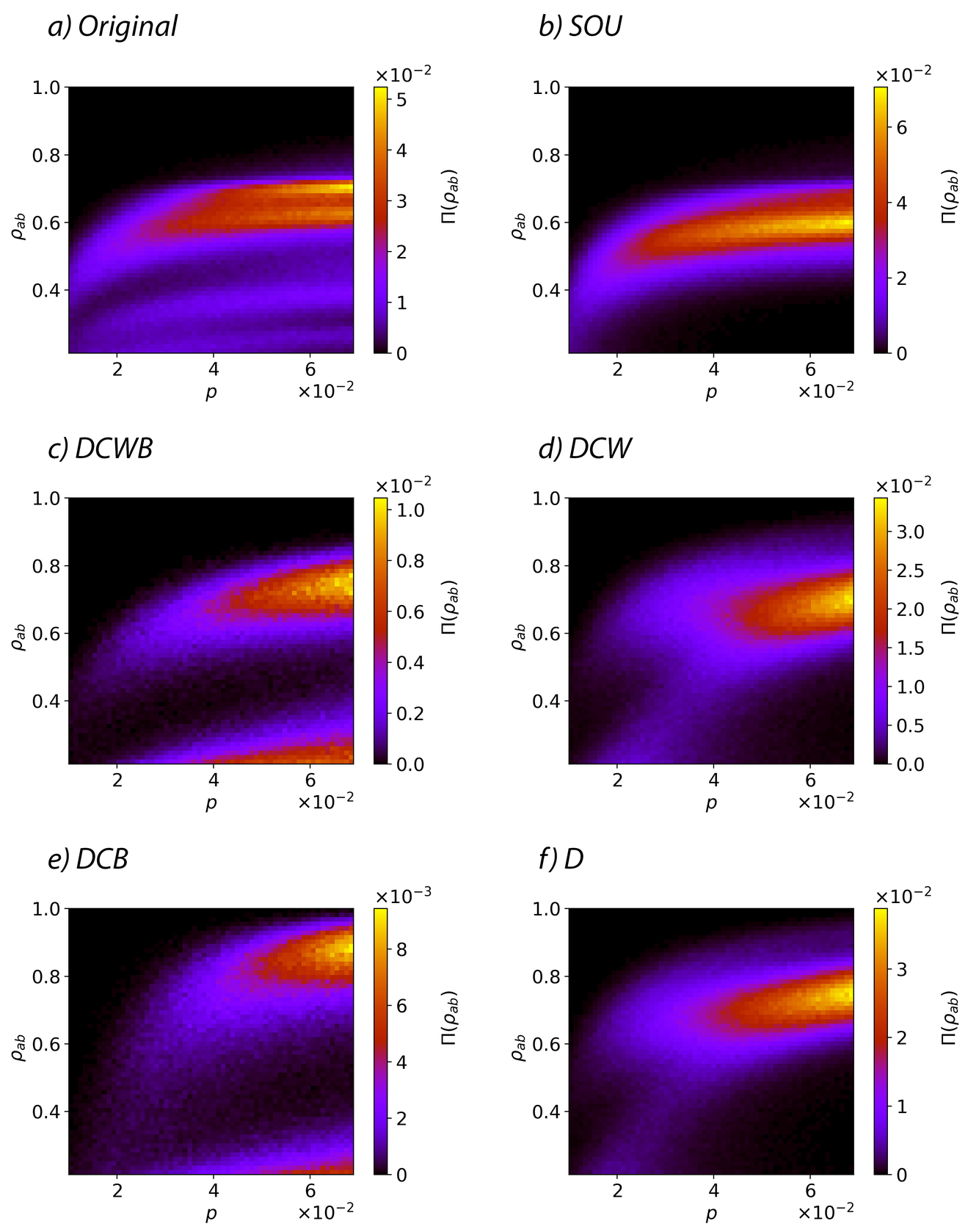}
		\caption{
			The results of \textbf{coinfective SIR-SIR ($\mathbf{q=1}$) simulation on the hospital network}, the x axis is the control parameter $p$, the y axis is the size of the final doubly infected agents ($ab$), and the color code denotes the percentage of realization with the specific value of $ab$.
			Please note that the coloring scale is different on each graph, to clarify the discrepancy between different regions of each graph.
		}
		\label{hospital-c}
	\end{center}
\end{figure*}

\begin{figure*}[]
	\begin{center}
		\includegraphics[width=\linewidth]{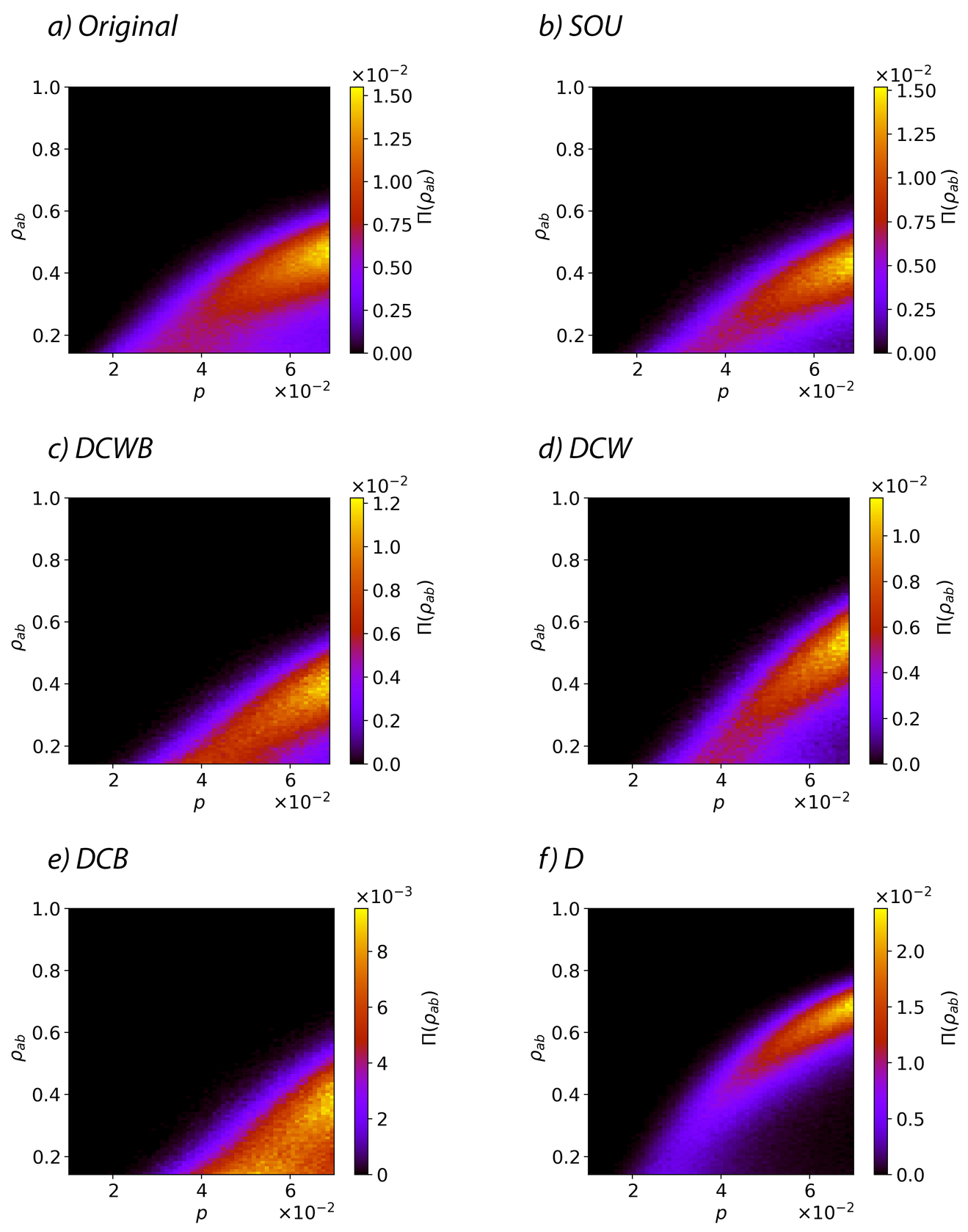}
		\caption{
			The results of \textbf{independent SIR-SIR infection ($\mathbf{q=p}$) simulation on the conference network}, the x axis is the control parameter $p$, the y axis is the size of the final doubly infected agents ($ab$), and the color code denotes the percentage of realization with the specific value of $ab$.
			Please note that the coloring scale is different on each graph, to clarify the discrepancy between different regions of each graph.
		}
		\label{conf-nc}
	\end{center}
\end{figure*}

\begin{figure*}[]
	\begin{center}
		\includegraphics[width=\linewidth]{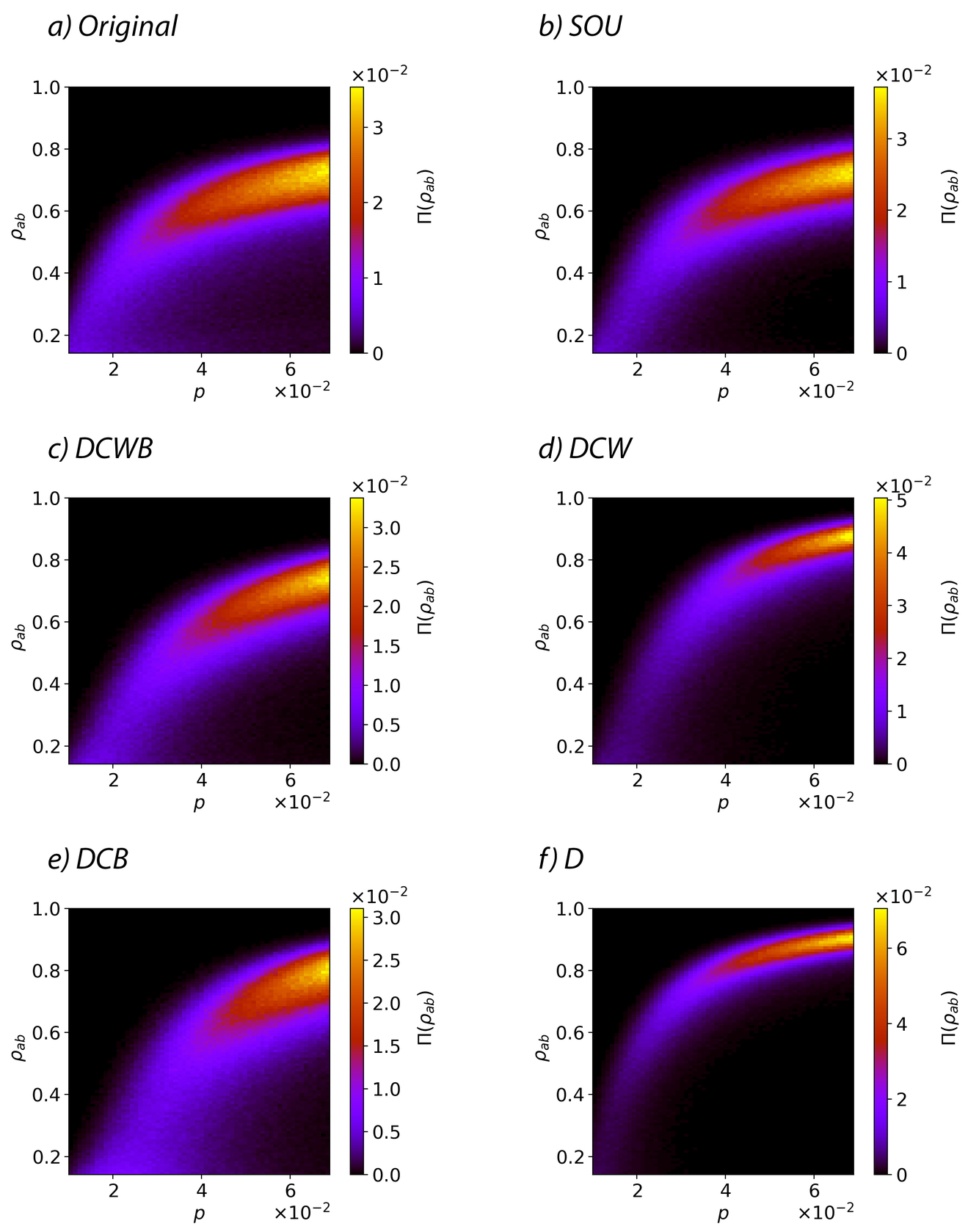}
		\caption{
			The results of \textbf{coinfective SIR-SIR ($\mathbf{q=1}$) simulation on the conference network}, the x axis is the control parameter $p$, the y axis is the size of the final doubly infected agents ($ab$), and the color code denotes the percentage of realization with the specific value of $ab$. Please note that the coloring scale is different on each graph, to clarify the discrepancy between different regions of each graph.
		}
		\label{conf-c}
	\end{center}
\end{figure*}

\begin{figure*}[]
	\begin{center}
		\includegraphics[width=\linewidth]{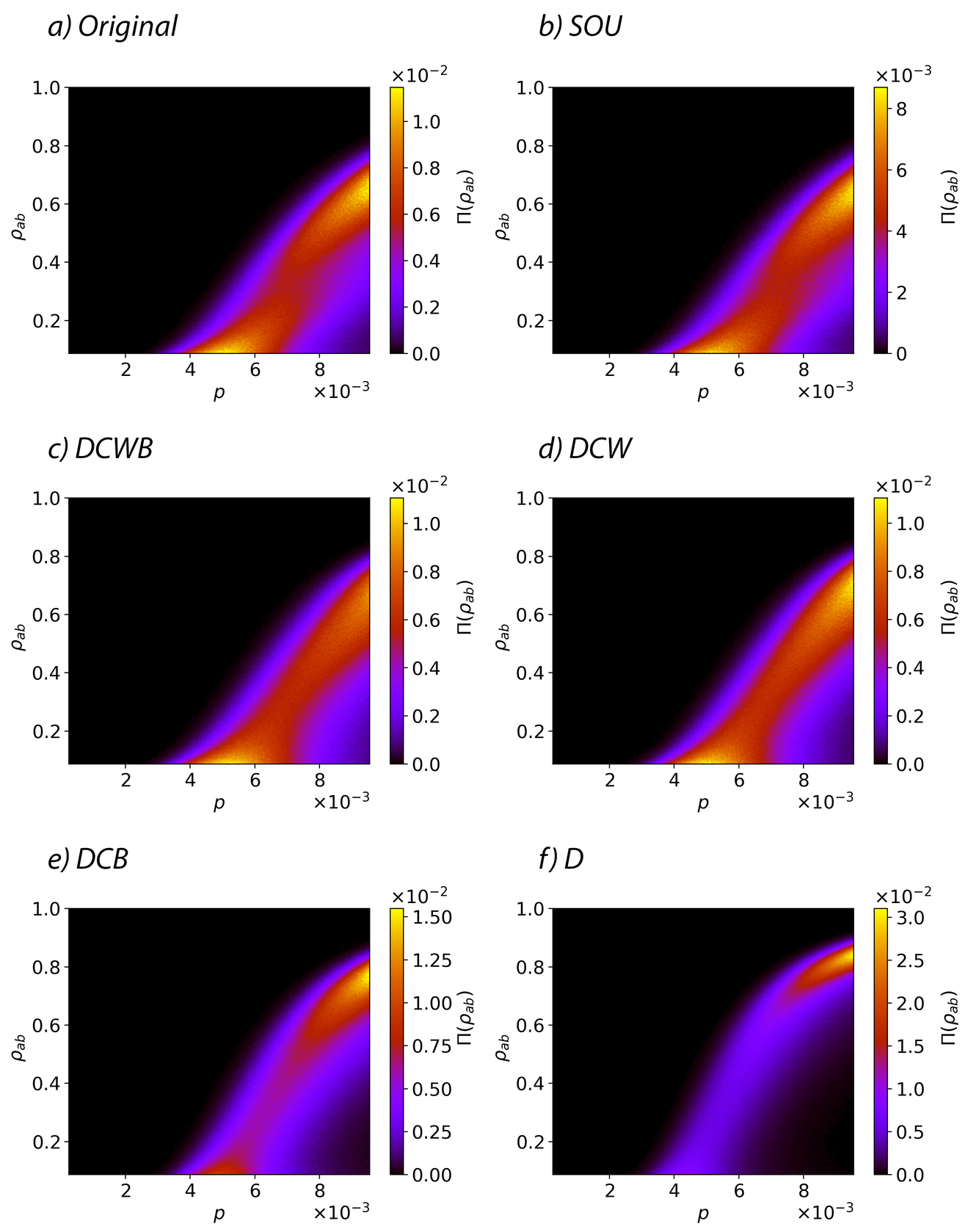}
		\caption{
			The results of \textbf{independent SIR-SIR infection ($\mathbf{q=p}$) simulation on the primary school network}, the x axis is the control parameter $p$, the y axis is the size of the final doubly infected agents ($ab$), and the color code denotes the percentage of realization with the specific value of $ab$.
			Please note that the coloring scale is different on each graph, to clarify the discrepancy between different regions of each graph.
		}
		\label{school-nc}
	\end{center}
\end{figure*}

\begin{figure*}[]
	\begin{center}
		\includegraphics[width=\linewidth]{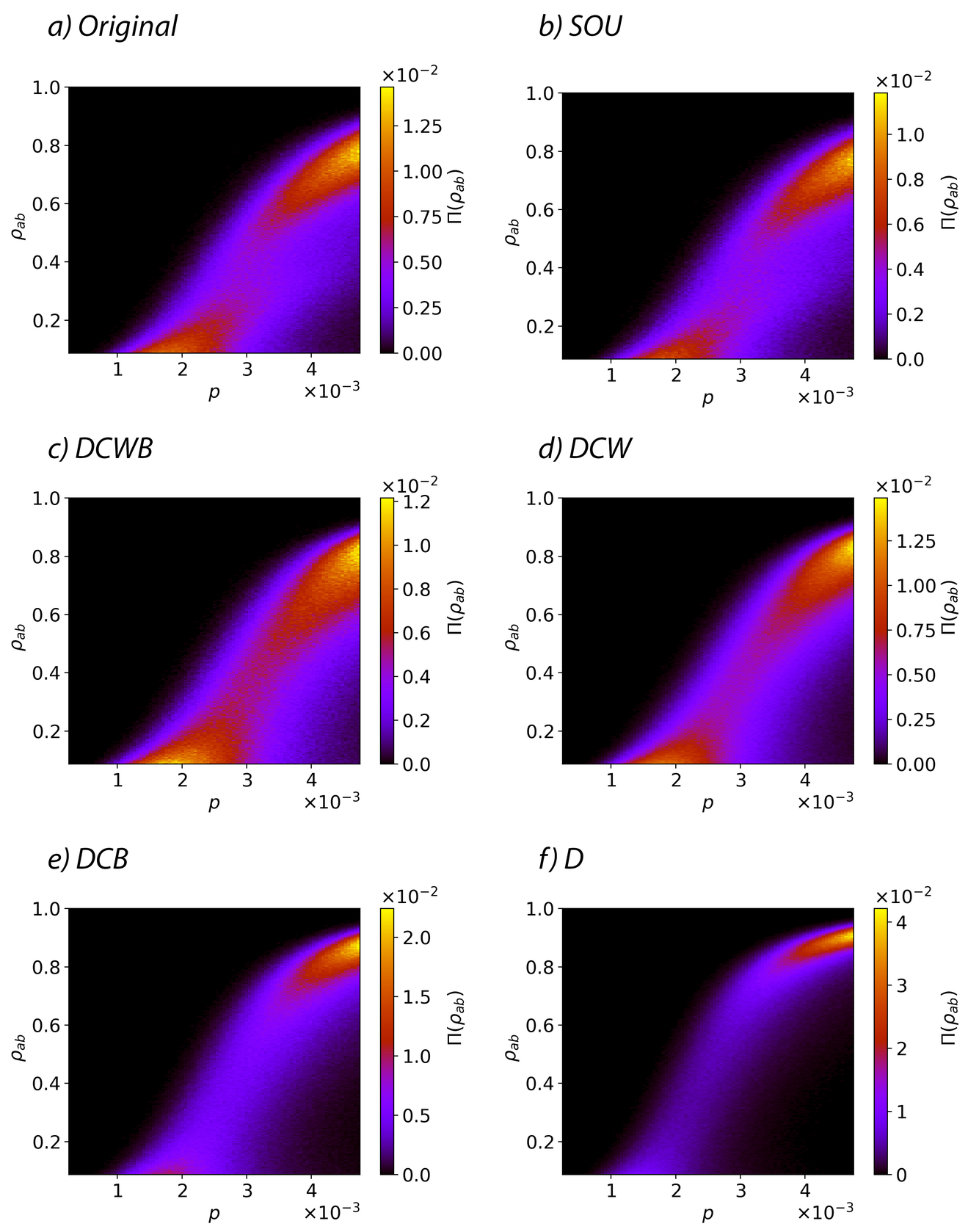}
		\caption{
			The results of \textbf{coinfective SIR-SIR ($\mathbf{q=1}$) simulation on the primary school network}, the x axis is the control parameter $p$, the y axis is the size of the final doubly infected agents ($ab$), and the color code denotes the percentage of realization with the specific value of $ab$.
			Please note that the coloring scale is different on each graph, to clarify the discrepancy between different regions of each graph.
		}
		\label{school-c}
	\end{center}
\end{figure*}

\clearpage
\bibliographystyle{abbrvnat}


\end{document}